\documentclass[11pt,notitlepage]{article}

\usepackage[utf8]{inputenc}
\usepackage{amsmath}
\usepackage{amsfonts}
\usepackage{graphics}
\usepackage{graphicx}
\usepackage{amssymb}
\usepackage{epigraph}
\usepackage{latexsym}
\usepackage{ifthen}
\usepackage{setspace}
\usepackage{epsfig}
\usepackage{multirow}
\usepackage{booktabs}
\usepackage{array}
\usepackage{float}
\usepackage{verbatim}
\usepackage{rotating}
\usepackage{caption}
\usepackage{subcaption}
\usepackage{comment}
\usepackage{stata}
\usepackage{bm}
\usepackage{tikz}
\usetikzlibrary{arrows,decorations.pathmorphing,backgrounds,positioning,fit,petri,matrix,trees,shadows}
\usepackage{authblk}
\usepackage[numbers]{natbib}
\usepackage{geometry}
\title{Assessing and relaxing the Markov assumption in the illness-death model}
\author[1, $\star$]{Jonathan Broomfield}
\author[2]{Caroline E. Weibull}
\author[3]{Michael J. Crowther}
\affil[1]{Biostatistics Research Group, Department of Health Sciences, University of Leicester, Leicester, UK}
\affil[2]{Clinical Epidemiology Division, Department of Medicine Solna, Karolinska Institutet, Stockholm, Sweden}
\affil[3]{Department of Medical Epidemiology and Biostatistics, Karolinska Institutet, Stockholm, Sweden}
\affil[$\star$]{jb781@le.ac.uk}
\date{}

\begin{document}

\maketitle

\begin{abstract}
Multi-state survival analysis considers several potential events of interest along a disease pathway. Such analyses are crucial to model complex patient trajectories and are increasingly being used in epidemiological and health economic settings. Multi-state models often make the Markov assumption, whereby an individual's future trajectory is dependent only upon their present state, not their past. In reality, there may be transitional dependence upon either previous events and/or more than one timescale, for example time since entry to the current or previous state(s). The aim of this study was to develop an illness-death Weibull model allowing for multiple timescales to impact the future risk of death. Following this, we evaluated the performance of the multiple timescale model against a Markov illness-death model in a set of plausible simulation scenarios when the Markov assumption was violated. Guided by a study in breast cancer, data were simulated from Weibull baseline distributions, with hazard functions dependent on single and multiple timescales. Markov and non-Markov models were fitted to account for/ignore the underlying data structure. Ignoring the presence of multiple timescales led to bias in underlying transition rates between states and associated covariate effects, while transition probabilities and lengths of stay were fairly robustly estimated. Further work may be needed to evaluate different estimands or more complex multi-state models. Software implementations in Stata are also described for simulating and estimating multiple timescale multi-state models.
\end{abstract}

\section{Background}
Multi-state models extend survival analysis to settings in which the risk of experiencing a particular event is in part determined by the occurrence of one or more intermediate events. This is a vital part of disease modelling, since most if not all diseases are far too complex to be characterised only by those that are alive and those that are dead. Multi-state models enable the entire patient pathway to be modelled, rather than focusing on one transition at a time. This allows more complex, time-varying relationships between covariates or intermediate events to be modelled. It also provides a framework for economic decision models in health technology assessment (HTA).

A common assumption in multi-state models is to constraint transitions to be dependent upon a subject's current state, and not on their disease history. This is known as the Markov assumption, and under it model fitting and predicting is straightforward. The assumption is rarely evaluated or relaxed, since accessible methods are limited. This paper illustrates the importance of considering the impact of the Markov assumption, and when to instead model transitions in multi-state models using more than one timescale.

\begin{figure}[h] \centering
\includegraphics[scale=0.7]{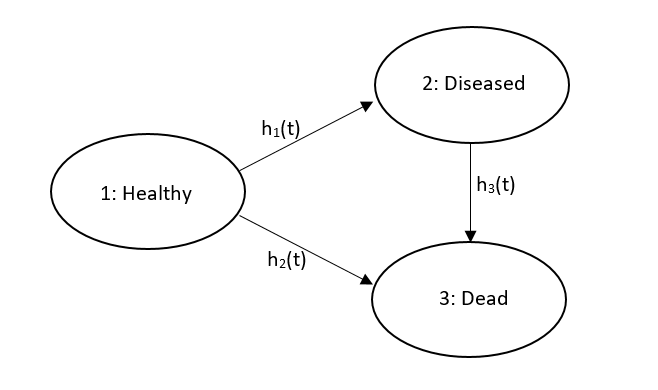}
\caption{Illness-death model, with transitions $h_i(t)$ between states.}
\label{fig:illdeath}
\end{figure}

A basic example of a multi-state model is the illness-death model, which has 3 states: typically these could represent healthy, ill and dead subjects. Subjects that are healthy would be expected to have a different risk of death to subjects that are ill. The structure of an illness-death model is shown in Figure \ref{fig:illdeath}.
 
In the illness-death model, while the transition rates $h_1(t)$ and $h_2(t)$ can depend upon additional timescales such as calendar time or patient age, they are generally dependent upon fewer timescales than $h_3(t)$. This is because all patients are typically in state 1 at origin (the beginning of a study period), and so transitions out of this state will not depend on previous state occupancies. Imposing the Markov assumption on the rate $h_3(t)$ does not allow any dependence on the time at which a patient became ill, denoted $r$. There are many epidemiological settings in which this will not be a plausible assumption. One possibility is to relax the model to a ``semi-Markov'', or clock-reset, model whereby the timescale for $h_3$ is time since entry to the ill state, $t - r$. However, the transition from ill to dead may well depend upon more timescales than just time since entry.

Putter et al provide a sound introduction to multi-state survival analysis through competing risks and the illness-death model\cite{put07}. They provide analytic solutions to calculating transition probabilities in Markov and semi-Markov settings from Cox transition models. The literature has also investigated the functional form that multiple timescale dependencies can take, and how to correctly adjust for it. Meier-Hirmer and Schumacher\cite{mei13} consider modelling hazards $h_1(t)$ and $h_2(t)$ both jointly and separately. They propose three approaches; the first is to model transitions from state 1 to 3 and state 2 to 3 in one function and assume a time-varying covariate effect indicating whether or not the subject is in state 2. This relationship can vary depending on the clinical or statistical importance, and could be proportional or dependent upon time at or time since entry to the second state. Transitions from state 1 to 2 are modelled separately. The second and third approaches models all three transitions separately, and so do not constrain baseline hazards to be the same for transitions 1 $\rightarrow$ 3 and 2 $\rightarrow$ 3. They differ by implementing different timescales to transitions from 2 $\rightarrow$ 3 - either time since origin (clock-forward) or time since entry to state 2 (clock-reset). The authors suggest the use of fractional polynomials to investigate the shape of the timescales, choosing between models via the AIC, and suggested that non-Markov structures should be appropriately accounted for.

Tassistro et al\cite{tas19} develop an algorithm for determining whether the hazard $h_3(t)$ in an illness-death structure is governed by multiple timescales from sub-sample models. Based on simulation results, they suggested that first fitting clock-forward and clock-reset models to the sample of subjects that move from state 2 to state 3 is sufficient for determining whether a Markov, semi-Markov or non-Markov approach should be adopted based on observed dependence on multiple timescales from the sub-sample.

Iacobelli and Carstensen\cite{iac13} suggest nesting single timescale models within more complex models that model multiple timescales linearly and using splines to assess the statistical importance of the more complex model structures. They propose plotting mortality rates over time since origin, a mortality rate ratio of a subject with a given entry time to state 2 against a subject that does not enter state 2 over time since entry to state 2, and another rate ratio comparing different relapse times to investigate the three timescales that this paper has mentioned. Plotting these rates and ratios on a common, logarithmic scale will make comparison and evaluation simpler.

Datta and Satten have shown that state occupancy probabilities calculated from non-Markovian data can be consistently estimated under the Markov assumption using non-parametric Nelson-Aalen estimators\cite{dat02}. However, parametric model estimations have not been investigated; little assessment has also gone into the robustness of other estimands, such as length of stay (LOS), under multiple timescale dependence. It is unknown, for instance, whether ignoring more complex model structures is exacerbated or partly alleviated when baseline transition rates are higher or lower, or when covariates are adjusted for in the analysis. Most pertinently, estimands of multi-state survival analysis need to be assessed for robustness to possible departures from the true underlying data structure. Studying this in more detail will have implications on the methodology of multi-state analysis as well as its application in areas such as HTA.

Still, transition probabilities are inconsistently calculated, as a literature review by Olariu et al\cite{ola17} found. Different papers propose non-parametric methods\cite{bri03}, numeric integration\cite{hin13} and simulation\cite{cro17}. Simulation is the most flexible approach developed to date, allowing transition-specific distributions with application to estimating clinically useful measures of effect differences\cite{cro17} as well as providing a straightforward approach to quantifying the uncertainty of the probabilities.


In this study, we developed a parametric multistate multiple-timescale model using Weibull baseline transitions between states. Model parameter estimates were obtained through maximum likelihood estimation, conducted using numerical integration. Predictions were obtained using a simulation approach. We investigated the robustness of parameter estimates, transition probabilities and LOSs under the Markov by simulating non-Markov scenarios and comparing these models to ones that accounted for the underlying data structure. Section \ref{sec:methods} describes the development of this model through transition-specific models, as well as the estimands assessed by simulation. Section \ref{sec:simulation} details the data-generation, ADEMP structure\cite{mor19} and results of the simulation while Section \ref{sec:demo} demonstrates the methods on a breast cancer dataset with an illness-death structure. We conclude the paper with a discussion in Section \ref{sec:discussion}.
 

\section{Methodology}
\label{sec:methods}

\subsection{The illness death model}

A multi-state model is formally defined as a stochastic process $ Y(t) \subset \lbrace 1, 2, 3...\rbrace$ where at time $t \geq 0$, the value of $Y$ is given by the state the patient is currently in. The transition matrix in a multi-state model defines which state transitions patients are allowed to make and which they are not. It provides the rate at which these occur as a function of some timescale(s). In the illness-death model, then when the final transition $h_3$ is dependent upon some timescale(s) relating to $t$ and/or $r$, the transition matrix $M$ is defined as:

\begin{equation*}
M = 
\begin{pmatrix}
0 & h_1(t) & h_2(t) \\
0 & 0 & h_3(t,r) \\
0 & 0 & 0
\end{pmatrix}
\end{equation*}
\\

Under the clock-forward approach, there is no dependence upon $r$.

Another common assumption in survival analysis is that proportional hazards are assumed between different subject demographics. The effect of a covariate $X_1$ on the hazard function $h_k(t), k \in \lbrace 1, 2, 3 \rbrace$ is assumed to be multiplicative, in the form of a hazard ratio $\exp(X_1\beta_{k,1})$. The baseline hazard functions $h_{0,k}(t)$ in this paper are assumed to follow Weibull distributions. This allows for a relatively flexible baseline transition rate in each case, and for proportional hazards between covariate levels. The Weibull survival distribution is also an easily invertible function and so provides a standard choice in simulation studies for generating survival times\cite{ben08}. The baseline Weibull model assumed in all clock-forward models across each transition is given below; $z_k(t,r)$ represents the potential effect of further timescales as functions of time since origin and time at entry to state 2. Note that $z_k(t,r) = 0$ for $k = 1,2$ since these transitions are out of state 1 and so will not depend on $r$.

\begin{align}
\label{eq:msmt_haz}
h_k(X|t,r) &= h_{0,k}(t) \exp(X\beta_k + z_k(t,r)) \\
&= \lambda_k \gamma_k t^{\gamma_k-1} \exp(X\beta_k + z_i(t,r)) \nonumber
\end{align}

This multiple timescale Weibull model can be applied to any general transition in a multi-state model. Baseline shape and scale parameters $\lambda_k$ and $\gamma_k$, covariate parameters $\beta_k$ and any parameters in the multiple timescale component $z_k(\cdot)$ are all estimated by maximising the likelihood of the hazard function in equation \ref{eq:msmt_haz} via numerical integration. Other baseline distributions, such as (piecewise) exponential or spline functions, can easily be substituted in.

\subsection{Multiple timescales} \label{sec:timescales}
This paper considers three cases of multiple timescale dependence, as functions of time since origin, $t$, and time at entry to state 2, $r$. In each instance, the only transition that will depend upon more than just $t$ will be $h_3(\cdot)$. The three cases are defined by the following transition functions (for some functions $f$ and $g$).

\begin{enumerate}
    \item $z_3(t,r) = f(r)$
    \item $z_3(t,r) = f(t-r)$
    \item $z_3(t,r) = f(r)g(t-r)$
\end{enumerate}

These all depend in the baseline case on time since origin, and are thus extensions of clock-forward models. The first case includes dependence on time at entry to state 2, the second on time since entry and the third on both time at entry and time since entry (with the introduction of a second function $g$ to indicate that the relationships of $r$ and $t-r$ on the hazard $h_3$ do not need to be the same.

Model transitions are estimated by maximising the likelihood of the $k^{th}$ transition intensity $h_k(t) = h_{0,k}(t)\exp(X\beta_k + z_k(t,r))$. In this paper, only linear functions of multiple timescales are considered, i.e. $f(x) = g(x) \delta x$ in cases 1-3 above. For example, this leads to the following full hazard function in case 3.

\begin{equation} \label{eq:haz3}
h_3(X|t,r) = \lambda_3 \gamma_3 t^{\gamma_3-1} \exp(X\beta_3 + \delta_1 r + \delta_2 (t-r)) 
\end{equation}

The $i^{th}$ patient's contribution to the likelihood (and log-likelihood) function for the final transition is derived accordingly from Equation \ref{eq:haz3}; $d_i$ denotes the censoring indicator.

\begin{align}
\label{eq:like_haz3}
L_{3,i} &= S_3(t_i,r_i) \cdot h_3(t_i,r_i)^{d_i} \nonumber \\
&= \exp(-\lambda_3 t_i^{\gamma_3}\exp(X_i\beta_3 + \delta_1 r_i + \delta_2 (t_i - r_i))) \cdot (\lambda_3 \gamma_3 t_i^{\gamma_3-1} \exp(X_i\beta_3 + \delta_1 r_i + \delta_2 (t_i - r_i)))^{d_i} \nonumber \\
\rightarrow log(L_{3,i}) := LL_{3,i} &= -\lambda_3 t_i^{\gamma_3}\exp(X_i\beta_3 + \delta_1 r_i + \delta_2 (t_i - r_i)) + d_i*(log(\lambda_3 \gamma_3 t_i^{\gamma_3-1}) + X_i\beta_3 + \delta_1 r_i + \delta_2 (t_i - r_i))
\end{align}

The overall log-likelihood for the third transition is then calculated by summing the term in Equation \ref{eq:like_haz3} over all observations $i$. This is estimated by numerical integration, which this paper demonstrates to be an accurate procedure when correctly accounting for model structure. These methods extend beyond illness-death settings and allow users to flexibly model any transition in a multi-state model and include dependence upon multiple timescales, and further to investigate the functional form of these relationships.

\subsection{Prediction}
Transition-specific models in multi-state survival analysis are interpreted in the same way as in a standard survival setting (which equates to a two-state model. For instance, each $h_k(t)$  in Figure \ref{fig:illdeath} corresponds to a hazard function estimating the rate at which subjects move from one state to another.

\subsubsection{Transition Probabilities}
An important application of multi-state survival is the estimation of transition probabilities. These predict the probability of a patient moving from one state to another over a given time interval $[s,t]$. One especially useful set of probabilities are the probabilities of a patient being in any state at time $t$ given they started in state 1 at origin ($s = 0$). These special cases of transition probabilities are sometimes referred to as state occupancy probabilities.

This paper investigated these particular transition probabilities for a patient starting in state 1 at time 0. In a conventional illness-death model, under the Markov assumption/ clock-forward approach, there are analytic solutions to estimating the 3 transition probabilities, given below\citep{put07}. Here, $p_{ij}(u,t)$ denotes the probability of a subject being in state $j$ at time $t$ given they were in state $i$ at time $u$.

\begin{align*}
p_{11}(u,t) &= \frac{S_1(t)}{S_1(u)} \,; \\
p_{12}(u,t) &= \frac{\int_u^t h_1(r)S_1(r)p_{22}(r,t)\,dr}{S_1(u)} \,; \\
p_{13}(u,t) &= 1 - p_{11}(u,t) - p_{12}(u,t)                                                      
\end{align*}
\\

To see how these equations are reached, first note that all probabilities refer to transitions between times $u$ and $t$ - they are thus conditional on survival up to time $u$. As such, the transitions $p_{11}(u,t)$ and $p_{12}(u,t)$ are divided by the survival function to time $u$ to incorporate this condition. The transition $p_{11}(u,t)$ is the simplest case to consider; it is simply the probability of remaining in state 1 from time $u$ to time $t$; thus, it is the survival function to time $t$ divided by the function to time $u$.

Next, to understand the transition probability $p_{12}(u,t)$, first consider the integrand term $h_1(r)S_1(r)$. In order to move from state 1 to state 2 before time $t$, there must be some time $r$ at which the subject move to state 2, expressed as the survival function of not moving out of state 1, $S_1(r)$, multiplied by the transition specific hazard of moving from state 1 to state 2, $h_1(r)$.

Then, once in state 2, the probability of remaining in state 2 is simply denoted $p_{22}(r,t)$. Note the $r$ in the expression rather than $u$ in the original expression since this transition time will vary from patient to patient and all possible transition times must be considered. This probability is comparable with $p_{11}(u,t)$ - it is the probability of remaining in state 2 from time $r$ to time $t$, and so can be expressed as the survival function from state 2 to 3, $S_3$, evaluated at time $t$ divided by $S_3$ evaluated at time $r$. Integrating from $u$ to $t$ covers all possible transition times $r$. The final transition probability, $p_{13}(u,t)$, can simply be calculated as a complement of the other transition probabilities; if a subject is not in state 1 or state 2 at time $t$ then, in an illness-death model, they must be in state 3.

In the clock-reset setting, the transition probability $p_{22}(r,t)$ is no longer calculated in the same way as $p_{22}(u,t)$, since a different timescale is being used (time since entry to state 2 - the clock is reset), and thus $p_{12}(u,t)$ is instead calculated accordingly\cite{put07}:

\begin{equation*}
p_{12}(u,t) = \frac{\int_u^t h_1(r)S_1(r)\frac{S_3(t-r)}{S_3(u-r)}\,dr}{S_1(u)}
\end{equation*}
\\

When $u = 0$, as in this paper, the denominator $S_2(u-r)$ is set to 1, since survival from state 2 at time 0 is guaranteed, and so the probability simplifies.

The clock-forward multiple timescale probabilities are more complicated, since $p_{22}$ depends upon $t$ as well as $r$. This relationship varies with the chosen approach to modelling the timescale (see Section \ref{sec:timescales}).

Since each transition will be constructed from a Weibull distribution, and thus the hazard function $h_i(t)$ will have baseline parameters $\lambda_i$ and $\gamma_i$ and effects $\beta_i$ of the covariates $X$, these transition probabilities can be expressed in exact formulae. The clock-forward, single timescale scenario is given below.

\begin{align*}
p_{11}(0,t) &= \exp(-\lambda_1 t^{\gamma_1} \exp(X\beta_1) -\lambda_2 t^{\gamma_2} \exp(\beta_2))\,; \\
p_{12}(0,t) &= \int_0^t \lambda_1 \gamma_1 r^{\gamma_1 -1} \exp(X\beta_1) \exp(-\lambda_1 r^{\gamma_1} \exp(X\beta_1) - \lambda_2 r^{\gamma_2} \exp(X\beta_2))\frac{\exp(-\lambda_3 t^{\gamma_3} \exp(X\beta_3))}{\exp(-\lambda_3 r^{\gamma_3} \exp(X\beta_3))} \,dr \,; \\
p_{13}(0,t) &= 1 - p_{11}(0,t) - p_{12}(0,t)                                                      
\end{align*}
\\

While these equations can sometimes be solved analytically, the expressions quickly become complicated where additional timescales are added to the model. This paper used the \textit{predictms} prediction command within \textit{multistate} to calculate transition probabilities by numerical integration, while standard errors were calculated using the delta method\cite{cro17}.

\subsubsection{Length of Stay}
LOS is a commonly used estimand in HTA, and is defined as the expected time period that a subject spends in any or each state within a defined timeframe, $[u_1, u_2]$. LOS is defined mathematically as such:

\begin{equation*}
L_j = \int_{u_1}^{u_2} p_{ij}(t) \, dt = \int_{u_1}^{u_2} S_j(t) \, dt
\end{equation*}
\\

Here, $L_j$ gives the expected length of stay in state $j$ for a subject starting in state $i$ in the interval $[u_1, u_2]$. Since HTA multi-state analysis evaluates the cost-effectiveness of a treatment as a disease progresses, the starting state $i$ often defaults to state 1, the starting time $u_1$ to 0 and the end time $u_2$ to death. This gives the total LOS estimate for time spent in state $j$ (assuming $j$ is not an absorbing state) as:

\begin{equation*}
L_{TOT,j} = \int_0^{\infty} p_{1j}(t) \, dt = \int_0^{\infty} S_j(t) \, dt
\end{equation*}

These estimates can be calculated using numeric integration, while uncertainty is often quantified using the delta method.

The LOS in states 1 and 2 under a clock-forward, single timescale model with a Weibull baseline hazard function, is given below.

\begin{equation*}
L_1 = \int_{u_1}^{u_2} \exp(-\lambda_1 t^{\gamma_1} \exp(X\beta_1) - \lambda_2 t^{\gamma_2} \exp(X\beta_2)) \, dt
\end{equation*}

\begin{align}
L_2 &= \int_{u_1}^{u_2} \int_0^t (\lambda_1\gamma_1 r^{\gamma_1 - 1} \exp(\beta_1)  + \lambda_2 \gamma_2 r^{\gamma_2 - 1} \exp(X\beta_2) ) \exp(-\lambda_1 r^{\gamma_1} \exp(X\beta_1) -\lambda_2 r^{\gamma_2} \exp(X\beta_2)) \nonumber  \\
&\times \frac{\exp(-\lambda_3 t^{\gamma_3} \exp(X\beta_3) )}{\exp(-\lambda_3 r^{\gamma_3} \exp(X\beta_3))} \, dr \, dt \nonumber
\end{align}

The LOS in state 3 can be calculated as the length of time between $u_1$ and $u_2$ not spent in states 1 or 2. To see this, note the following.

\begin{align*}
S_3(t) &= 1 - S_1(t) - S_2(t) \\
\rightarrow L_3 &= \int_{u_1}^{u_2} (1 - S_1(t) - S_2(t)) \, dt = u_2 - u_1 - L_1 - L_2
\end{align*}
\\

Since state 3 is an absorbing state, the quantity $L_3$ only has practical use in a finite sense, as total LOS in this state will be infinite. The limit of $L_3$ as the upper limit $u_2 \rightarrow \infty$ is also $\infty$.

\section{Simulation}
\label{sec:simulation}
In order to assess the impact of the Markov assumption in multi-state models, survival data were simulated that correspond to the three multiple timescale cases discussed in Section \ref{sec:timescales}. As such, the ``true'' form of the transition hazards will be known, and so therefore will estimands of interest such as parameter estimates, transition probabilities and lengths of stay. Note that while parameter estimates will be exactly known, transition probabilities and lengths of stay will only be known within a small tolerance, since the true values of these estimands must be calculated using the simulation approach of \textit{predictms}\citep{cro17}. Comparisons were made by fitting separate models that correct adjust for and that violate the underlying data structure.

Survival times were simulated, where possible, by inverting survival functions. This was implemented using \textit{survsim}\citep{cro11}, utilising the following relationship between a survival time $T_k$ and a Weibull survival function\cite{ben08}. Below, $U$ denotes a random draw from a standard uniform distribution while $k = 1,2$.

\begin{align*}
S(T_k) &= \exp(-\lambda_k T_k^{\gamma_k} \exp(X\beta_k)) = U \\
\rightarrow T_k &= \left(\frac{-\log((U)}{\lambda_k \exp(X\beta_k)}\right)^{\frac{1}{\gamma_k}}
\end{align*}
\\

For the transitions from state 2 to 3, the multiple timescale scenarios have more complicated baseline hazard functions. When simulating from the model including time at entry to state 2, $r$, as well as time since origin, $t$ (case one of the multiple timescale settings), the hazard is still invertible and so survival times were generated in a similar way to the single timescale setting. Note that these survival times were only simulated for patients whose simulated transition time to state 2 $T_1$ was smaller than their transition time to state 3 $T_2$. For these patients, $r$ was equal to their simulated $T_1$ value.

\begin{align*}
\frac{S(T_3)}{S(r)} &= \frac{\exp(-\lambda_3 T_3^{\gamma_3} \exp(X\beta_3 + \delta_1 r))}{\exp(-\lambda r^{\gamma_3} \exp(X\beta_3))} = U \\
\rightarrow T_3 &= \left(\frac{-\log(U)}{\lambda_3 \exp(X\beta_3 + \delta_1 r)} + r^{\gamma_3}\right)^{\frac{1}{\gamma_3}}
\end{align*}
\\

When simulating from models including time since entry, $t - r$ (cases 2 and 3 shown sequentially below) the hazard function is non-invertible in closed form. Iterative root finding was used to randomly generate survival times incorporating both timescale dependencies and delayed entry.

\begin{equation*}
\frac{S(T_3)}{S(r)} = \frac{\exp(-\lambda_3 T_3^{\gamma_3} \exp(X\beta_3 + \delta_2 (T_3-r)))}{\exp(-\lambda_3 r^{\gamma_1} \exp(X\beta_3))} = U
\end{equation*}

\begin{equation*}
\frac{S(T_3)}{S(r)} = \frac{\exp(-\lambda_3 T_3^{\gamma_3} \exp(X\beta_3 + \delta_1 r + \delta_2 (T_3-r)))}{\exp(-\lambda_3 r^{\gamma_1} \exp(X\beta_3))} = U
\end{equation*}
\\

\subsection{ADEMP Structure}
The simulation study followed the ADEMP protocol detailed by Morris et al\cite{mor19}. This provides clarity and ensures rigorous evaluation of methods by defining the aims, data-generating mechanism, estimands, methods and performance measures of the study. The ADEMP structure of this paper is as follows:

\begin{enumerate}
\item Aims: to assess the validity and impact of the Markov assumption under multiple timescale dependencies.
\item Data-generating mechanism: Weibull baseline hazard functions with additional timescale functions, simulated using \textit{survsim}. See below for more details.
\item Estimands: Transition parameters, and transition probabilities and LOS estimates each at 5 years.
\item Methods: Fitting transition specific models assuming Weibull baseline distributions and obtaining subsequent predictions of transition probabilities and LOS via \textit{predictms}. For each multiple timescale scenario, two models were fitted to the final transition; one using the multiple timescale hazard function developed in Equation \ref{eq:haz3} and one assuming a basic Weibull model with no additional timescales (i.e. Markov).
\item Performance measures: bias, coverage and their standard errors\cite{mor19}.
\end{enumerate}

Monte Carlo standard errors are used to quantify the uncertainty of performance measures such as bias and coverage. Assuming a desired coverage of 95\% and maximum Monte Carlo standard error (MCSE) of 1 requires a value of $n_{sim}$ of 475, while an MCSE of 0.75 gives $n_{sim}$ = 844\cite{mor19}. To be conservative, $n_{sim}$ was set to 1,000 for all scenarios in this paper, ensuring suitably low MCSEs.

Simulations were run 1,000 times on 2,000 simulated patients, approximately consistent with the sample size of the guiding breast cancer dataset. Two covariates were included in data-generation, representing a continuous variable ($X_1$, ``age'') and a binary variable ($X_2$, ``treatment''). For baseline Weibull distributions, shape and scale parameters of $\lambda_i = 0.1$ and $\gamma_i = 1.3$, and covariate effects $\beta_{1,i}/\beta_{2,i}$ of 0.01/0.5, were used throughout. When included in the model, the effect sizes of multiple timescales $\delta_i$ were set to 0.1.

\subsection{Results}

\begin{figure}[b!]
    \centering
    \begin{subfigure}[t]{0.5\textwidth}
        \centering
        \includegraphics[width=8cm]{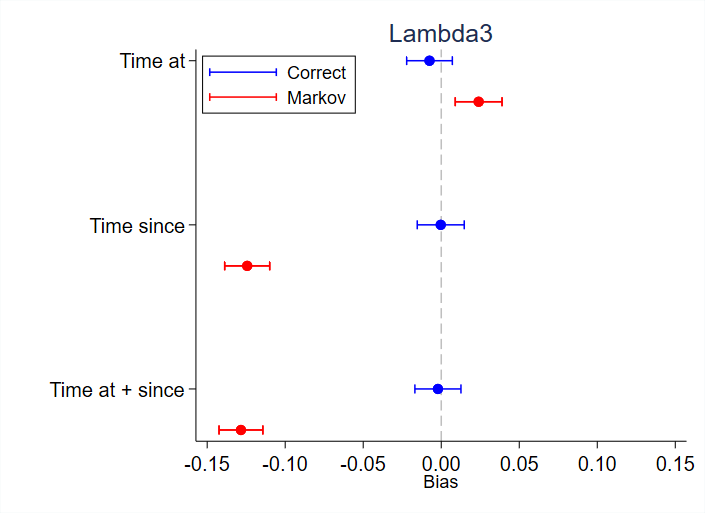}
        \caption{\normalsize{Scale $\lambda_3$}} \label{lam3_bias}
    \end{subfigure}%
    ~
    \begin{subfigure}[t]{0.5\textwidth}
        \centering
        \includegraphics[width=8cm]{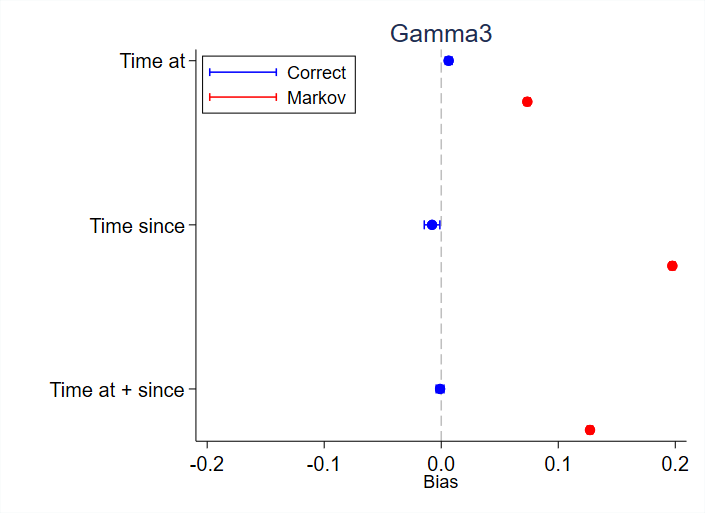}
        \caption{\normalsize{Shape $\gamma_3$}} \label{gam3_bias}
    \end{subfigure}
    
    \begin{subfigure}[t]{0.5\textwidth}
        \centering
        \includegraphics[width=8cm]{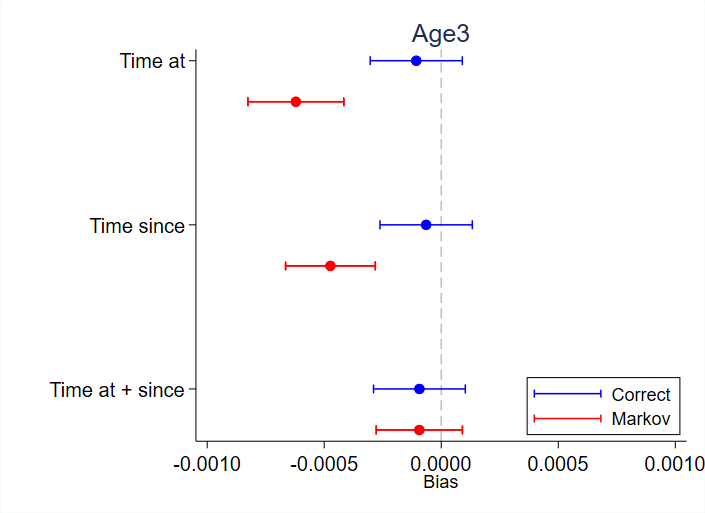}
        \caption{\normalsize{Continuous $\beta_{1,3}$}} \label{age3_bias}
    \end{subfigure}%
    ~
    \begin{subfigure}[t]{0.5\textwidth}
        \centering
        \includegraphics[width=8cm]{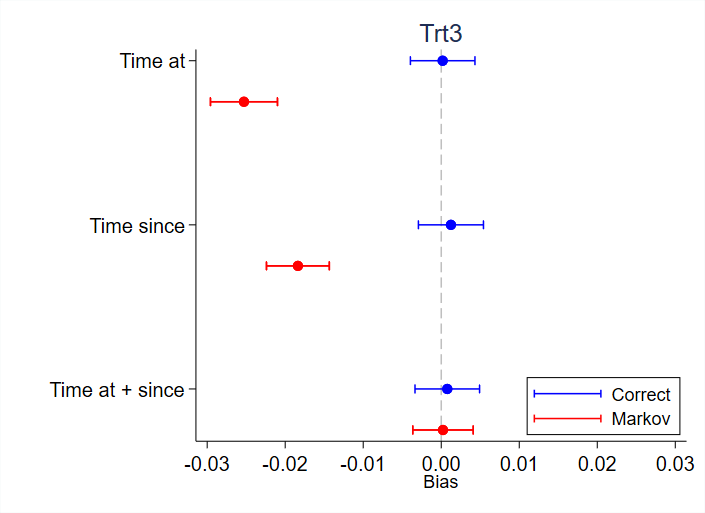}
        \caption{\normalsize{Binary $\beta_{2,3}$}} \label{trt3_bias}
    \end{subfigure}
    
    \caption{Bias of final transition parameters under multiple timescale data simulation.} \label{fig:param_bias}
\end{figure}

\begin{figure}[h!]
    \centering
    \begin{subfigure}[t]{0.5\textwidth}
        \centering
        \includegraphics[width=8cm]{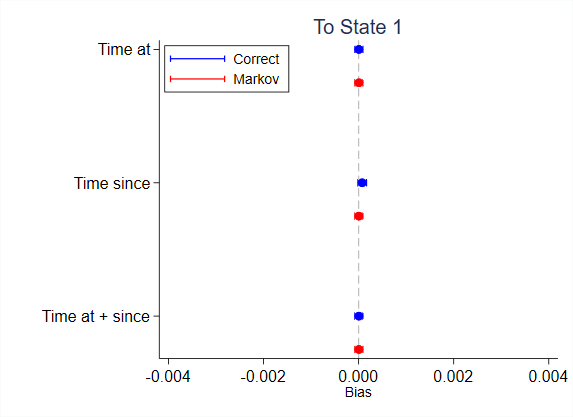}
        \caption{Transition probability to state 1} \label{p1_bias}
    \end{subfigure}%
    ~
    \begin{subfigure}[t]{0.5\textwidth}
        \centering
        \includegraphics[width=8cm]{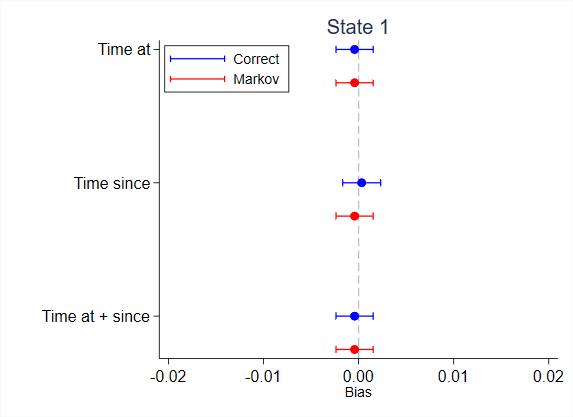}
        \caption{LOS estimate in state 1} \label{los1_bias}
    \end{subfigure}
    
    \begin{subfigure}[t]{0.5\textwidth}
        \centering
        \includegraphics[width=8cm]{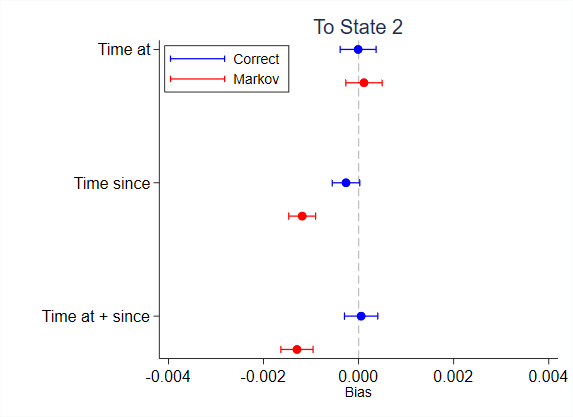}
        \caption{Transition probability to state 2} \label{p2_bias}
    \end{subfigure}%
    ~
    \begin{subfigure}[t]{0.5\textwidth}
        \centering
        \includegraphics[width=8cm]{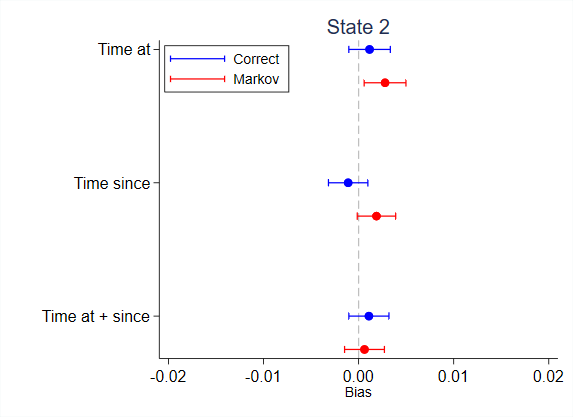}
        \caption{LOS estimate in state 2} \label{los2_bias}
    \end{subfigure}
    
    \begin{subfigure}[t]{0.5\textwidth}
        \centering
        \includegraphics[width=8cm]{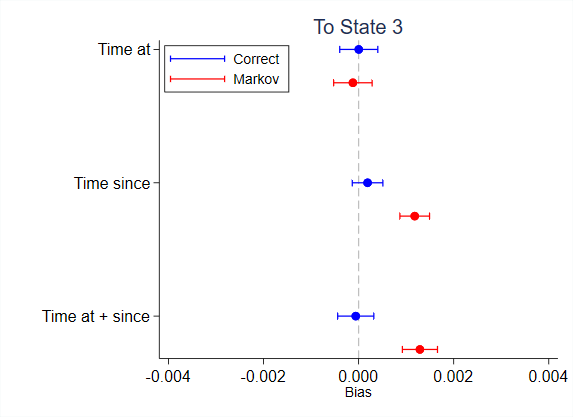}
        \caption{Transition probability to state 3} \label{p3_bias}
    \end{subfigure}%
    ~
    \begin{subfigure}[t]{0.5\textwidth}
        \centering
        \includegraphics[width=8cm]{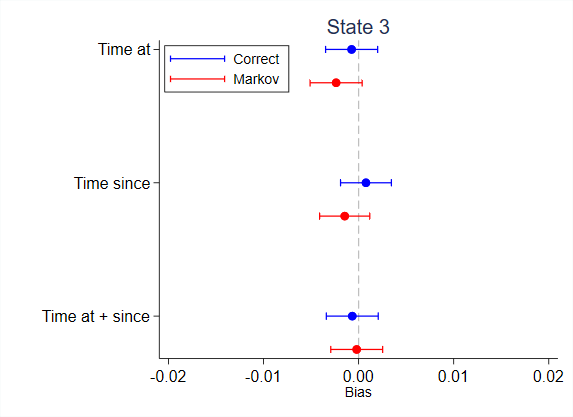}
        \caption{LOS estimate in state 3} \label{los3_bias}
    \end{subfigure}
    \captionsetup{justification=centering}
    \caption{Bias of transition probabilities and LOS estimates at 5 years under multiple timescale data simulation.} \label{fig:translos_bias}
\end{figure}

The results are presented assessing in turn parameter estimates and transition probabilities/LOS estimates.

Overall, estimands were accurately reproduced when correct time dependence was accounted for. Since it is becoming increasingly more straightforward to fit multiple timescale models, this performance highlights the importance of considering transitional dependence upon more than one timescale when fitting multi-state models. It demonstrates the simulation based approach towards multi-state estimation and calculations of transition probabilities/LOS through \textit{merlin} and \textit{predictms} to be an accurate yet efficient one\cite{cro17, cro18}.

\subsubsection{Parameter estimates}

The estimand most severely affected by a failure to adjust for multiple timescales was parameter estimates. Only parameters from the final transition, state 2 $\rightarrow$ 3, were adversely affected since this was the only transition to be incorrectly modelled. Figure \ref{fig:param_bias} shows the bias of $\lambda_3, \gamma_3, \beta_{1,3}$ and $\beta_{2,3}$ from each of the transition functions discussed, with confidence intervals (CIs) calculated using MCSEs.

The underlying Weibull distributions were inaccurately estimated when models were misspecified. In particular, the scale parameter $\lambda_3$ was subject to negative bias, underestimating the true mortality rate of ill patients. The shape parameter $\gamma_3$ was positively biased, which may well relate to the observed negative bias of $\lambda_3$, with one compensating for the other. Covariate effects were also inaccurately estimated in relative terms, although the absolute impact of this was minimal. Coverage, shown in Appendix Figure \ref{fig:param_cover}, was very poor for the shape parameter $\gamma_3$ when the multiple timescale dependence was ignored.

\subsubsection{Transition probabilities and LOS estimates}

Transition probabilities and LOS estimates were very robust to departures from the true model structure. As the transitions were estimated from origin (time 0), this is consistent with non-parametric investigation by Datta and Satten\cite{dat02}, and so appropriates the use of Markov models in certain HTA settings where outcomes and decisions are governed entirely by these estimands. LOS estimates were also robustly estimated. This is possibly due to the direct relationship between the two, although no previous work has investigated LOS as an estimand in non-Markov settings. Figure \ref{fig:translos_bias} shows the bias each model exhibited when calculating transition probabilities from state 1 to each of the three states, and the predicted LOS in each state, 5 years after origin. The data-generation was non-Markov, and correct and Markov models are contrasted in the figure.

The apparent relative bias observed in models is in fact very minimal, with the magnitude of bias being very low. Coverage, shown in Appendix Figure \ref{fig:translos_cover}, was fairly good in all cases, highlighting the suitability of the delta method to calculate standard errors in \textit{predictms}.

\section{Rotterdam dataset}
\label{sec:demo}
The simulations, in order to be biologically plausible\cite{cro13}, were guided by the Rotterdam breast cancer dataset. This contains time-to-event survival data for 2,982 women until relapse and/or death. The methods and software are now illustrated on the dataset.

\subsection{Demographics}
\begin{figure}[h] \centering
\includegraphics[scale=0.7]{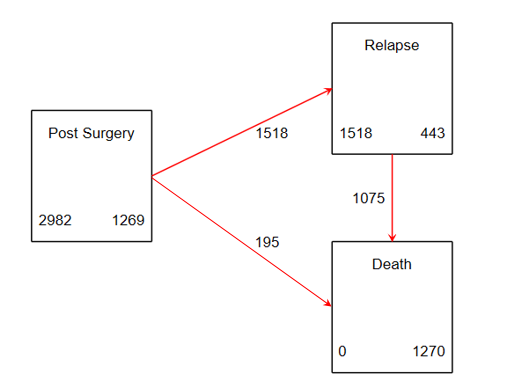}
\caption{Illness-death structure of the Rotterdam dataset.}
\label{fig:illdeath_rott}
\end{figure}

The dataset has a 3 state, illness-death structure; patients are either in post-surgery (state 1), in relapse (state 2) or dead (state 3), as shown in Figure \ref{fig:illdeath_rott}. Numbers in the bottom left and right of boxes respectively give the number at risk at the start and end of follow-up. The baseline demographics of the 2,982 patients are given in Table \ref{tab:demo}.

\begin{table}[h!]
\centering
\caption{Baseline demographics of the Rotterdam dataset.}
\label{tab:demo}
\begin{tabular}{ccclcc}
\hline
\textbf{Variable}                                                                & \textbf{Level}     & \textbf{N (\%)} &  & \textbf{Variable}                                                                  & \textbf{Mean (sd)}                                                           \\ \hline
Menopause                                                                        & Pre                & 1312 (44.0)     &  & \multirow{2}{*}{\begin{tabular}[c]{@{}c@{}}Age\\ (years)\end{tabular}}             & \multirow{2}{*}{\begin{tabular}[c]{@{}c@{}}55.06\\ (13.0)\end{tabular}}      \\
                                                                                 & Post               & 1670 (56.0)     &  &                                                                                    &                                                                              \\
                                                                                 &                    &                 &  &                                                                                    &                                                                              \\
\multirow{2}{*}{\begin{tabular}[c]{@{}c@{}}Differentiation\\ Grade\end{tabular}} & 2                  & 794 (26.6)      &  & \multirow{2}{*}{\begin{tabular}[c]{@{}c@{}}Surgery Year\\ (years)\end{tabular}} & \multirow{2}{*}{\begin{tabular}[c]{@{}c@{}}1988\\ (3.0)\end{tabular}}  \\
                                                                                 & 3                  & 2188 (73.4)     &  &                                                                                    &                                                                              \\
                                                                                 &                    &                 &  &                                                                                    &                                                                             
                                                                                 \\
\multirow{2}{*}{\begin{tabular}[c]{@{}c@{}}Hormonal\\ Therapy\end{tabular}} & Yes                  & 339 (11.4)      &  & \multirow{2}{*}{\begin{tabular}[c]{@{}c@{}}No. Nodes\\ (\#)\end{tabular}} & \multirow{2}{*}{\begin{tabular}[c]{@{}c@{}}2.71\\ (4.4)\end{tabular}}  \\
                                                                                 & No                  & 2643 (88.6)     &  &                                                                                    &                                                                              \\
                                                                                 &                    &                 &  &                                                                                    &                                                                               \\
Chemotherapy                                                                     & Yes                & 2402 (80.6)     &  & \multirow{2}{*}{\begin{tabular}[c]{@{}c@{}}PgR Level\\ (fmol/l)\end{tabular}}      & \multirow{2}{*}{\begin{tabular}[c]{@{}c@{}}161.83\\ (291.3)\end{tabular}}    \\
                                                                                 & No                 & 580 (19.4)      &  &                                                                                    &                                                                              \\
                                                                                 &                    &                 &  &                                                                                    &                                                                              \\
Tumour Size                                                                      & \textless{}= 20mm  & 1387 (46.5)     &  & \multirow{2}{*}{\begin{tabular}[c]{@{}c@{}}ER Level\\ (fmol/l)\end{tabular}}       & \multirow{2}{*}{\begin{tabular}[c]{@{}c@{}}166.59\\ \\ (272.5)\end{tabular}} \\
                                                                                 & 20-50mm            & 1291 (43.3)     &  &                                                                                    &                                                                              \\
                                                                                 & \textgreater{}50mm & 304 (10.2)      &  &                                                                                    &                                                                              \\ \hline
\end{tabular}
\end{table}

\subsection{Illustration of methods}
Model implementation is now demonstrated using \textit{merlin}. Full Stata output including tables of parameter estimates are in the Appendix Section \ref{sec:appB}. The data is first loaded, rescaled and inspected.

\begin{stlog}
*Load data
use rott2, clear

*Rescale times from months to years
replace rf = rf/12
replace os = os/12

*Inspect data
list pid rf rfi os osi if inlist(pid,1,1371), noobs sepby(pid)

  +-----------------------------------+
  |  pid    rf   rfi    os        osi |
  |-----------------------------------|
  |    1   4.9     0   4.9      alive |
  |-----------------------------------|
  | 1371   1.4     1   2.0   deceased |
  +-----------------------------------+
  
\end{stlog}

In order to estimate the three hazard function, variables must be created for each transition. This can be done using \textit{msset}, part of the \textit{multistate} package\cite{cro17}, which creates internal variables. \textit{msset} reshapes the data from wide to long format, although this is not necessary for fitting multi-state models with \textit{merlin}.

\begin{stlog}
*Reshape to obtain start and stop times
msset, id(pid) states(rfi osi) times(rf os)

*Re-inspect data
list pid rf rfi os osi if inlist(pid,1,1371), noobs nolab sepby(pid)

  +-----------------------------------+
  |  pid    rf   rfi    os        osi |
  |-----------------------------------|
  |    1   4.9     0   4.9      alive |
  |    1   4.9     0   4.9      alive |
  |-----------------------------------|
  | 1371   1.4     1   2.0   deceased |
  | 1371   1.4     1   2.0   deceased |
  | 1371   1.4     1   2.0   deceased |
  +-----------------------------------+

list pid _start _stop _from _to _status _trans if inlist(pid,1,1371), noobs sepby(pid)

  +---------------------------------------------------------------+
  |  pid      _start       _stop   _from   _to   _status   _trans |
  |---------------------------------------------------------------|
  |    1           0   4.9253936       1     2         0        1 |
  |    1           0   4.9253936       1     3         0        2 |
  |---------------------------------------------------------------|
  | 1371           0   1.3798767       1     2         1        1 |
  | 1371           0   1.3798767       1     3         0        2 |
  | 1371   1.3798767   2.0287473       2     3         1        3 |
  +---------------------------------------------------------------+

\end{stlog}

Each transition can now be modelled using \textit{merlin}. The first two transitions are the same for all single and multiple timescale models. Weibull baseline models with a variety of explanatory covariates are fitted\cite{cro17}, using \textit{stmerlin} since single timescales are used. One variable, tumour size, is split into two binary covariates. Model estimates are stored for later use.

\begin{stlog}
*Create indicator variables for tumour size
tab size, gen(sz)

Tumour size |      Freq.     Percent        Cum.
------------+-----------------------------------
    <=20 mm |      3,339       44.63       44.63
  >20-50mmm |      3,327       44.47       89.09
     >50 mm |        816       10.91      100.00
------------+-----------------------------------
      Total |      7,482      100.00

*stset the data for stmerlin commands
stset _stop, fail(_status)

*Model first transition (state 1 -> 2, same for all models)
stmerlin age sz2 sz3 nodes pr_1 hormon if _trans == 1, dist(weibull)
est store mod1

*Model second second transition (state 1 -> 3, same for all models)
stmerlin age sz2 sz3 nodes pr_1 hormon if _trans == 2,  dist(weibull)
est store mod2
\end{stlog}

Each model's hazard function for the final transition is now estimated. Numerical integration is used for models including time since entry to state 2, specified with the \textit{timevar} option, while multiple timescales are included in the model as a restricted cubic spline with 1 degree of freedom (i.e. linear) to aid predictions. Again, Weibull baseline models with a variety of explanatory covariates are fitted. Note a Royston-Parmar restricted cubic spline model with 1 degree of freedom is actually specified, since this aids with likelihood estimation in multiple timescale models and is statistically equivalent to a Weibull model\cite{rut15}. The \textit{noorthog} option provides baseline estimates that correspond to the shape and scale parameters $\lambda$ and $\gamma$

\begin{stlog}
*Fit final transition models assuming Weibull baseline
*First clock-forward (no multiple timescales)
merlin (_stop age sz2 sz3 nodes pr_1 hormon if _trans == 3, ///
    family(rp, df(1) fail(_status) ltruncated(_start) noorthog))
est store mod3_cf

*Second including time at entry
merlin (_stop age sz2 sz3 nodes pr_1 hormon rcs(_start, df(1)) if _trans == 3, ///
    family(rp, df(1) fail(_status) ltruncated(_start) noorthog))
est store mod3_ta

*Third including time since entry
merlin (_stop age sz2 sz3 nodes pr_1 hormon ///
    rcs(_stop, df(1) moffset(_start)) if _trans == 3, ///
    family(rp, df(1) fail(_status) ltruncated(_start) noorthog) timevar(_stop))
est store mod3_ts

*Fourth including time at and time since entry
merlin (_stop age sz2 sz3 nodes pr_1 hormon rcs(_start, df(1)) ///
    rcs(_stop, df(1) moffset(_start)) if _trans == 3, ///
    family(rp, df(1) fail(_status) ltruncated(_start) noorthog) timevar(_stop))
est store mod3_tas

\end{stlog}

Shape and scale parameters are compared in Table \ref{tab:demo_param}. 

\begin{table}[h]
\centering
\caption{Shape and scale parameters from final transition in Rotterdam dataset under each clock-forward model.}
\label{tab:demo_param}
\begin{tabular}{ccccc}
\hline
\textbf{Model}        & \textbf{Shape $\lambda_3$} & \textbf{$SE_{\lambda_3}$} & \textbf{Scale $\gamma_3$} & \textbf{$SE_{\gamma_3}$} \\ \hline
Clock-forward         & 0.552           & 0.20        & 0.668          & 0.05        \\
Time at entry         & 0.448           & 0.19        & 0.850          & 0.06        \\
Time since entry      & 0.723           & 0.27        & 0.501          & 0.10        \\
Time at + since entry & 0.515           & 0.24        & 0.743          & 0.11        \\ \hline
\end{tabular}
\end{table}

Transitional hazard dependence on time at entry to state 2 of approximately -0.1 on the log scale was observed. This corresponds to a 10\% decrease in the rate of transition to state 3 for each year later a patient moved to state 2. Dependence on time since entry was approximately 0.01, leading to a 1\% increase in the rate of transition for each year a patient remains in state 2. These are fairly mild effects of multiple timescale dependence, but the shape and scale parameters fairly considerably. In particular, the basic clock-forward model appears to overestimate $\lambda_3$ and underestimate $\gamma_3$ compared to the final model including all timescales, which should be the model used since there is evidence of multiple timescale dependence. Covariate effects were estimated quite consistently across each model. For a full comparison of each model's estimates, see Appendix Section \ref{sec:appB}.

\begin{figure}[h]
    \centering
    \begin{subfigure}[t]{0.5\textwidth}
        \centering
        \includegraphics[width=8cm]{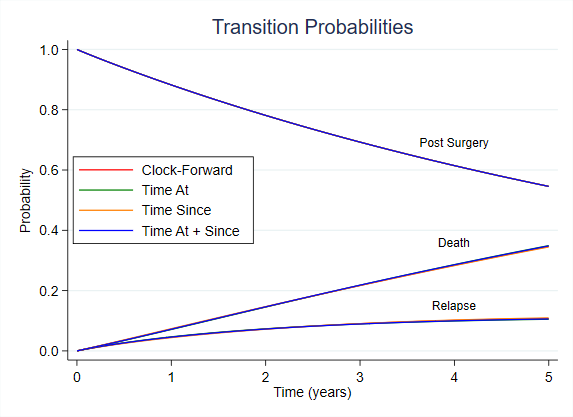}
    \end{subfigure}%
    ~
    \begin{subfigure}[t]{0.5\textwidth}
        \centering
        \includegraphics[width=8cm]{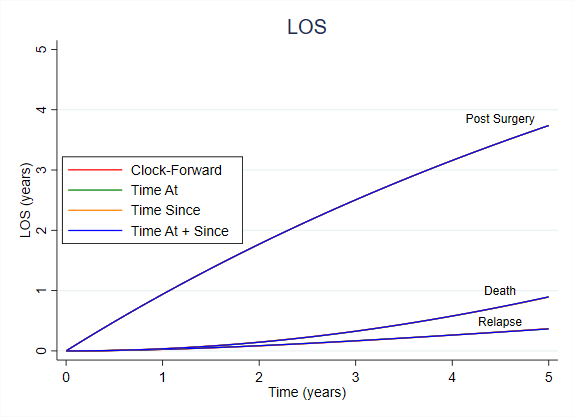}
    \end{subfigure}
    
\caption{Transition probabilities and LOS estimates from Rotterdam dataset under each model.} \label{fig:rott_trans_los_models}
\end{figure}

These models have been stored and so can be used to calculate transition probabilities and LOS estimates with \textit{predictms}. These estimates are compared at 5 years for each model in Figure \ref{fig:rott_trans_los_models}. Confidence intervals can be calculated using the delta method by specifying the \textit{ci} option. Predictions are made for a patient aged 60 with the smallest tumour size, 0 nodes, log(pr+1) = 1, and not on hormonal therapy.

\begin{stlog}
matrix tmat = (.,1,2 \textbackslash .,.,3 \textbackslash .,.,.)

*Compare models' transition probabilities and LOS estimates in each state
cap drop t
range t 0 5 101
foreach method in cf ta ts tas \{
    cap drop `method'_prob_*
	cap drop `method'_los_*
    predictms, transmatrix(tmat) models(mod1 mod2 mod3_`method') prob los ///
	timevar(t) at1(age 60 sz2 0 sz3 0 nodes 0 pr_1 1 hormon 0) ci
    rename _prob_* `method'_prob_*
	rename _los_* `method'_los_*
\}
\end{stlog}

Both transition probabilities and LOS estimates are consistently estimated regardless of assumptions about multiple timescales, confirming their robustness to non-Markov settings. Note that predictions for the post surgery state are identical in all cases since transitions out of this state are the same for all single and multiple timescale models.

\section{Discussion}
\label{sec:discussion}
In this paper we have developed a multiple timescale Weibull survival model to consider more complex state relationships in multi-state analysis. We demonstrate this method to be accurate in estimation and prediction. We further assess robustness of the Markov assumption in contexts where multiple timescale dependence is present, by simulating a variety of biologically plausible scenarios under Markovian and non-Markovian frameworks. Models were fitted to the data constructed under non-Markov scenarios that did and did not account for the true nature of the multiple timescale dependencies, and compared to each other and to models from Markov settings.

\subsection{Context of Results}
What is most apparent from the results for the multiple timescale component of the analysis is that when a Markov framework is constrained on data that are dependent not just on time since origin, as Markov models assume, but also on functions of time at or since entry to the previous state, parameter estimates are highly likely to be biased with low coverage. It is more likely that the underlying hazard of transition will be incorrectly estimated, but also possible that covariate effects will be inaccurate. While the results in this paper correspond only to one transition from a very basic illness-death model, they likely translate to similar impacts on more complex scenarios. For example, it would be expected that assuming a transition in a more complex multi-state model has no dependency on subject history, when in fact there is a relationship between the history and the future, would result in biased parameter estimates. More complex non-Markov relationships being ignored in this manner may well compound the effects of bias. Given the transitions in a multi-state model can be considered as three individual survival models, this study also underlines the importance of accounting for appropriate multiple timescales in general survival settings.

The impact of misspecifying the relationship between a subject's history in a multi-state framework and their future is contextually dependent. In settings where a small dependency of hazard of death on time at entry or time since exists, the bias is likely to be more affected by the sample sizes, with larger datasets reducing bias. For instance, failing to correctly model the time dependence in a model dependent upon time since entry to state 2 as well as since origin led to error in the baseline rate of transition $\lambda_3$. This corresponded to a baseline 87 deaths per 1,000 person years being estimated, compared to the known truth of 100. Such a large absolute error may have arisen due to the high true rate of death, and the large time dependence that was ignored. However, this is not an unreasonable or uncommon scenario, having being taken from a real-life dataset, and more ``mild'' circumstances would still likely lead to poor discrepancies in model estimations.

In all multiple timescale settings, the final rate parameter $\lambda_3$ was negatively biased when ignoring the true data structure, leading to consistent underestimation of the true rate of death from subjects in state 2. The bias of covariate effects varied but tended to be positively biased when the effect was positive, which would lead to estimating that covariates were more extreme than was true. Transition probabilities and LOS estimates were found to be robust to departures from Markov settings, which allows for the use of Markov models when interest lies solely in these estimands - however, other estimands are likely to depend on the transition parameters themselves, and so require appropriate modelling of multiple timescales.

Care must thus be taken to avoid exaggerating the effects of treatments in HTAs. In order to do so, multiple timescale models should be considered in all multi-state survival analysis. There should be thought given to the biological plausibility of a dependence on history of the future, which exists in many medical settings.

These results were fairly consistent to what was observed when applying the methods to the Rotterdam breast cancer dataset. The transition probabilities were measured from origin and so were equivalent to state occupancy probabilities. Parameter estimates varied but transition probabilities and LOS estimates were consistent across different models.

\subsection{Limitations and Future Work}
Three different estimands were assessed by two performance measures. This increased computation time, particularly as model complexity increased and hazard functions were no longer invertible. This is also the first paper to investigate the impact of non-Markov settings on LOS estimates. Further work could focus on a greater variation of parameter estimates, both individually and in combination with one another, to gain a greater understanding of the situations in which it is most crucial to account for the possible presence of dependence upon multiple timescales. Alternatively, the impact of sample size could be investigated further.

Additionally, all simulations generated data under an illness-death model, a basic example of a multi-state model, and assumed that all transitions had a baseline hazard function that followed a Weibull distribution. Other survival models could also have been used; more flexible baseline models, such as splines or user-defined hazard functions, would provide more flexibility. The model could also be developed to incorporate time dependent covariates that are not captured by state definitions.

\section{Conclusion}
This paper has highlighted the issues of misspecifying Markovian multi-state models to data that are non-Markovian in nature. Failure to capture the true time dependencies of data can lead to bias in terms of both underlying rates of transition between states and covariate effects on transition rates. The observed bias correlates with the magnitude of the time dependence that is being ignored. It appears that the tendency is for covariate effects to be inflated, and for transition rates to be underestimated. The results have been placed in the wider context of health technology assessment, which is largely reliant upon such survival models \cite{son93, bri98, wil17} to make decisions on new or existing treatments based on their cost and efficacy. It is essential to consider the possible violation of the Markov assumption that is made in many analyses to simplify models, and the presence of a dependence upon the history of the subject's future, to avoid making harmfully incorrect decisions in these assessments, particularly in settings where rates of transition between states are high.

\bibliographystyle{wileyj}
\bibliography{paper_arxiv.bbl}

\newpage

\appendix
\section{Supplementary Figures}\label{sec:appA}

Coverage of the transition parameters, produced using \textit{rsurvsim}\cite{gas18,whi10}. The blue bars indicate CIs containing the true parameter value and red bars show those that do not.

\begin{figure}[h!]
    \centering
    \begin{subfigure}[t]{0.5\textwidth}
        \centering
        \includegraphics[width=6cm]{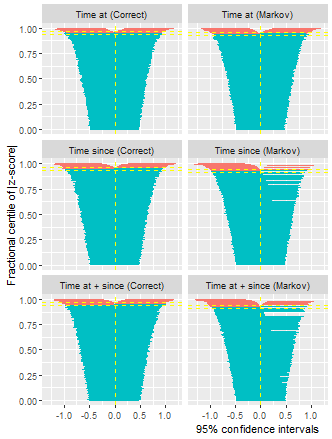}
        \caption{Scale $\lambda_3$} \label{lam3_cover}
    \end{subfigure}%
    ~
    \begin{subfigure}[t]{0.5\textwidth}
        \centering
        \includegraphics[width=6cm]{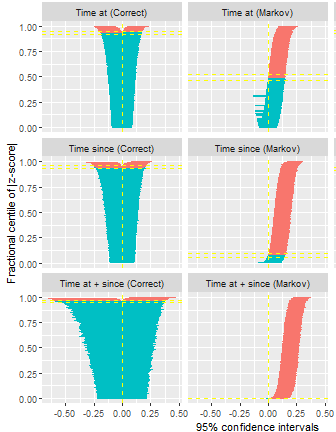}
        \caption{Shape $\gamma_3$} \label{gam3_cover}
    \end{subfigure}
    
    \begin{subfigure}[t]{0.5\textwidth}
        \centering
        \includegraphics[width=6cm]{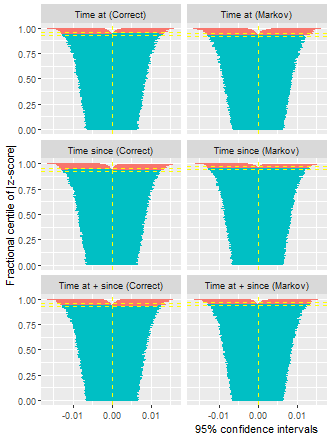}
        \caption{Continuous $\beta_{1,3}$} \label{age3_cover}
    \end{subfigure}%
    ~
    \begin{subfigure}[t]{0.5\textwidth}
        \centering
        \includegraphics[width=6cm]{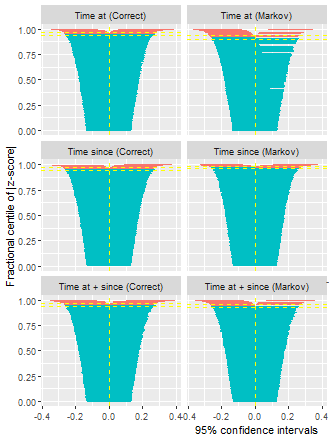}
        \caption{Binary $\beta_{2,3}$} \label{trt3_cover}
    \end{subfigure}
    
    \caption{Coverage of final transition parameters in multiple timescale settings under correct and Markov models.} \label{fig:param_cover}
\end{figure}

Coverage of the transition probabilities and LOS estimates at 5 years of follow up.

\begin{figure}[h!]
    \centering
    \begin{subfigure}[t]{0.5\textwidth}
        \centering
        \includegraphics[width=5cm]{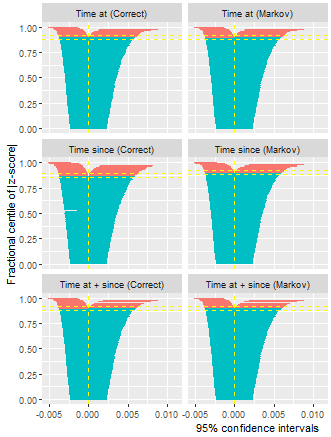}
        \caption{Transition probability to state 1} \label{p1_cover}
    \end{subfigure}%
    ~
    \begin{subfigure}[t]{0.5\textwidth}
        \centering
        \includegraphics[width=5cm]{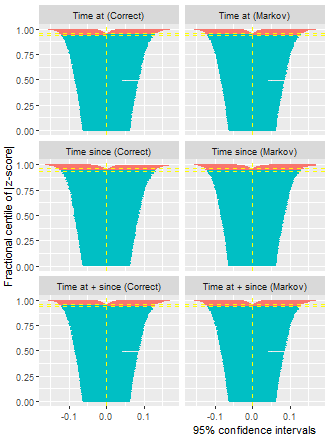}
        \caption{LOS estimate in state 1} \label{los1_cover}
    \end{subfigure}
    
    \begin{subfigure}[t]{0.5\textwidth}
        \centering
        \includegraphics[width=5cm]{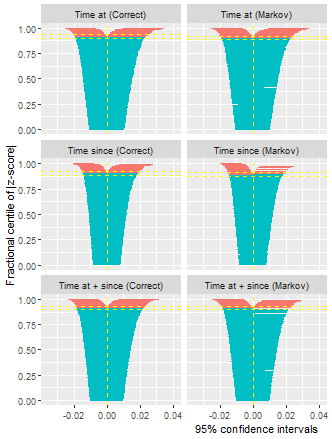}
        \caption{Transition probability to state 2} \label{p2_cover}
    \end{subfigure}%
    ~
    \begin{subfigure}[t]{0.5\textwidth}
        \centering
        \includegraphics[width=5cm]{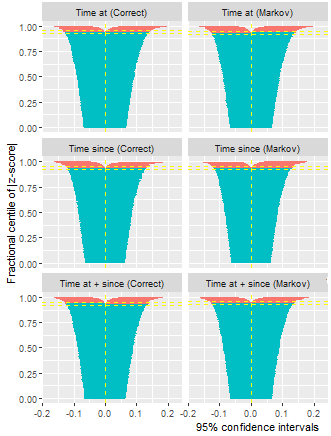}
        \caption{LOS estimate in state 2} \label{los2_cover}
    \end{subfigure}
    
    \begin{subfigure}[t]{0.5\textwidth}
        \centering
        \includegraphics[width=5cm]{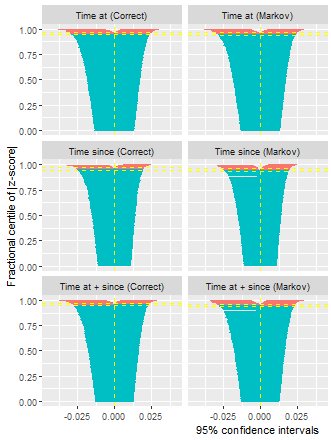}
        \caption{Transition probability to state 3} \label{p3_cover}
    \end{subfigure}%
    ~
    \begin{subfigure}[t]{0.5\textwidth}
        \centering
        \includegraphics[width=5cm]{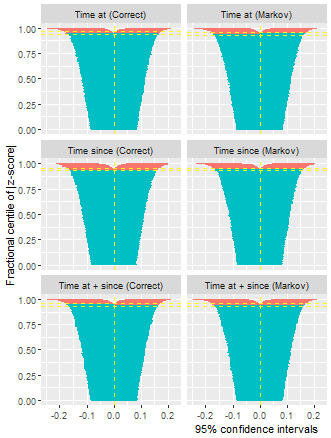}
        \caption{LOS estimate in state 3} \label{los3_cover}
    \end{subfigure}
    \captionsetup{justification=centering}
    \caption{Coverage of transition probabilities and LOS estimates at 5 years under multiple timescale data simulation.} \label{fig:translos_cover}
\end{figure}

\section{Illustration of methods - parameter estimates}\label{sec:appB}
Below is the full Stata output from Section \ref{sec:demo}.

\begin{stlog}
*Load data
use rott2, clear

*Rescale times from months to years
replace rf = rf/12
replace os = os/12

*Inspect data
list pid rf rfi os osi if inlist(pid,1,1371), noobs sepby(pid)

  +-----------------------------------+
  |  pid    rf   rfi    os        osi |
  |-----------------------------------|
  |    1   4.9     0   4.9      alive |
  |-----------------------------------|
  | 1371   1.4     1   2.0   deceased |
  +-----------------------------------+

*Reshape to obtain start and stop times
msset, id(pid) states(rfi osi) times(rf os)

*Re-inspect data
list pid rf rfi os osi if inlist(pid,1,1371), noobs nolab sepby(pid)

  +-----------------------------------+
  |  pid    rf   rfi    os        osi |
  |-----------------------------------|
  |    1   4.9     0   4.9      alive |
  |    1   4.9     0   4.9      alive |
  |-----------------------------------|
  | 1371   1.4     1   2.0   deceased |
  | 1371   1.4     1   2.0   deceased |
  | 1371   1.4     1   2.0   deceased |
  +-----------------------------------+

list pid _start _stop _from _to _status _trans if inlist(pid,1,1371), noobs sepby(pid)

  +---------------------------------------------------------------+
  |  pid      _start       _stop   _from   _to   _status   _trans |
  |---------------------------------------------------------------|
  |    1           0   4.9253936       1     2         0        1 |
  |    1           0   4.9253936       1     3         0        2 |
  |---------------------------------------------------------------|
  | 1371           0   1.3798767       1     2         1        1 |
  | 1371           0   1.3798767       1     3         0        2 |
  | 1371   1.3798767   2.0287473       2     3         1        3 |
  +---------------------------------------------------------------+

*Create indicator variabels for tumour size
tab size, gen(sz)

Tumour size |      Freq.     Percent        Cum.
------------+-----------------------------------
    <=20 mm |      3,339       44.63       44.63
  >20-50mmm |      3,327       44.47       89.09
     >50 mm |        816       10.91      100.00
------------+-----------------------------------
      Total |      7,482      100.00

*stset the data
stset _stop, fail(_status)

*Model first transition (state 1 -> 2, same for all models)
stmerlin age sz2 sz3 nodes pr_1 hormon if _trans == 1, dist(weibull)

Mixed effects regression model                  Number of obs     =      2,982
Log likelihood = -4962.3641
------------------------------------------------------------------------------
             |      Coef.   Std. Err.      z    P>|z|     [95\% Conf. Interval]
-------------+----------------------------------------------------------------
_stop:       |            
         age |  -.0062153   .0021012    -2.96   0.003    -.0103335    -.002097
         sz2 |   .3739369   .0580319     6.44   0.000     .2601965    .4876773
         sz3 |   .6799473   .0868836     7.83   0.000     .5096585    .8502361
       nodes |   .0811534   .0044792    18.12   0.000     .0723743    .0899326
        pr_1 |  -.0408656   .0115458    -3.54   0.000    -.0634949   -.0182362
      hormon |  -.0014572   .0821299    -0.02   0.986    -.1624288    .1595144
       _cons |  -.9625073   .1249859    -7.70   0.000    -1.207475   -.7175394
------------------------------------------------------------------------------

est store mod1

*Model second second transition (state 1 -> 3, same for all models)
stmerlin age sz2 sz3 nodes pr_1 hormon if _trans == 1, dist(weibull)

Mixed effects regression model                  Number of obs     =      2,982
Log likelihood =  -859.5294
------------------------------------------------------------------------------
             |      Coef.   Std. Err.      z    P>|z|     [95\% Conf. Interval]
-------------+----------------------------------------------------------------
_stop:       |            
         age |   .1250736   .0079699    15.69   0.000     .1094528    .1406943
         sz2 |   .1615512   .1614484     1.00   0.317    -.1548818    .4779842
         sz3 |   .4153081   .2332725     1.78   0.075    -.0418977    .8725138
       nodes |   .0439416   .0182545     2.41   0.016     .0081633    .0797198
        pr_1 |   .0223507   .0334238     0.67   0.504    -.0431588    .0878601
      hormon |  -.1399109   .2291894    -0.61   0.542     -.589114    .3092921
       _cons |  -11.64961   .5747208   -20.27   0.000    -12.77604   -10.52318
------------------------------------------------------------------------------

est store mod2

*Fit final transition models assuming Weibull baseline
*First clock-forward (no multiple timescales)
merlin (_stop age sz2 sz3 nodes pr_1 hormon if _trans == 3, ///
    family(rp, df(1) fail(_status) ltruncated(_start) noorthog))

Mixed effects regression model                  Number of obs     =      1,518
Log likelihood = -2385.5802
------------------------------------------------------------------------------
             |      Coef.   Std. Err.      z    P>|z|     [95\% Conf. Interval]
-------------+----------------------------------------------------------------
_stop:       |            
         age |   .0046747   .0024203     1.93   0.053     -.000069    .0094183
         sz2 |   .1697423     .07119     2.38   0.017     .0302125     .309272
         sz3 |   .3209264   .0994308     3.23   0.001     .1260456    .5158073
       nodes |   .0287836   .0057158     5.04   0.000     .0175809    .0399864
        pr_1 |  -.1033869   .0139645    -7.40   0.000    -.1307568   -.0760169
      hormon |   .0831566   .0967767     0.86   0.390    -.1065222    .2728354
       _cons |  -.5938329   .1993393    -2.98   0.003    -.9845308    -.203135
------------------------------------------------------------------------------

est store mod3_cf

*Second including time at entry
merlin (_stop age sz2 sz3 nodes pr_1 hormon rcs(_start, df(1)) if _trans == 3, ///
    family(rp, df(1) fail(_status) ltruncated(_start) noorthog))

Mixed effects regression model                  Number of obs     =      1,518
Log likelihood = -2374.5794
------------------------------------------------------------------------------
             |      Coef.   Std. Err.      z    P>|z|     [95\% Conf. Interval]
-------------+----------------------------------------------------------------
_stop:       |            
         age |   .0044778   .0024114     1.86   0.063    -.0002485    .0092041
         sz2 |   .1469844   .0712838     2.06   0.039     .0072708    .2866981
         sz3 |   .2893874   .0994476     2.91   0.004     .0944737    .4843011
       nodes |   .0294698   .0057547     5.12   0.000     .0181908    .0407488
        pr_1 |  -.1042988   .0139716    -7.47   0.000    -.1316826   -.0769149
      hormon |   .1022188   .0969888     1.05   0.292    -.0878757    .2923134
       rcs() |  -.0941164   .0204642    -4.60   0.000    -.1342255   -.0540073
       _cons |  -.8030618   .1943143    -4.13   0.000    -1.183911   -.4222128
------------------------------------------------------------------------------

est store mod3_ta

*Third including time since entry
merlin (_stop age sz2 sz3 nodes pr_1 hormon ///
    rcs(_stop, df(1) moffset(_start)) if _trans == 3, ///
    family(rp, df(1) fail(_status) ltruncated(_start) noorthog) timevar(_stop))

Mixed effects regression model                  Number of obs     =      1,518
Log likelihood = -2383.5035
------------------------------------------------------------------------------
             |      Coef.   Std. Err.      z    P>|z|     [95\% Conf. Interval]
-------------+----------------------------------------------------------------
_stop:       |            
         age |   .0047541   .0024224     1.96   0.050     6.16e-06     .009502
         sz2 |   .1697969   .0711876     2.39   0.017     .0302718     .309322
         sz3 |   .3206118   .0994385     3.22   0.001     .1257159    .5155077
       nodes |   .0288266   .0057216     5.04   0.000     .0176123    .0400408
        pr_1 |  -.1036385   .0139482    -7.43   0.000    -.1309765   -.0763006
      hormon |    .095345   .0970606     0.98   0.326    -.0948903    .2855803
       rcs() |   .0228007   .0112042     2.04   0.042     .0008409    .0447605
       _cons |  -.3237874   .2671735    -1.21   0.226    -.8474379    .1998631
------------------------------------------------------------------------------

est store mod3_ts

*Fourth including time at and time since entry
merlin (_stop age sz2 sz3 nodes pr_1 hormon rcs(_start, df(1)) ///
    rcs(_stop, df(1) moffset(_start)) if _trans == 3, ///
    family(rp, df(1) fail(_status) ltruncated(_start) noorthog) timevar(_stop))

Mixed effects regression model                  Number of obs     =      1,518
Log likelihood = -2373.9383
------------------------------------------------------------------------------
             |      Coef.   Std. Err.      z    P>|z|     [95\% Conf. Interval]
-------------+----------------------------------------------------------------
_stop:       |            
         age |   .0045211    .002413     1.87   0.061    -.0002082    .0092505
         sz2 |   .1469913    .071306     2.06   0.039     .0072342    .2867484
         sz3 |   .2884076   .0995155     2.90   0.004     .0933609    .4834543
       nodes |   .0293371   .0057601     5.09   0.000     .0180475    .0406267
        pr_1 |   -.103976   .0139687    -7.44   0.000     -.131354   -.0765979
      hormon |   .1072623   .0971374     1.10   0.269    -.0831236    .2976481
       rcs() |  -.0869102   .0209092    -4.16   0.000    -.1278915   -.0459289
       rcs() |   .0142904   .0127305     1.12   0.262     -.010661    .0392417
       _cons |  -.6635607   .2392022    -2.77   0.006    -1.132388   -.1947331
------------------------------------------------------------------------------

est store mod3_tas

matrix tmat = (.,1,2\textbackslash.,.,3\textbackslash.,.,.)

*Compare models' transition probabilities and LOS estimates in each state
cap drop t
range t 0 5 101
foreach method in cf ta ts tas \{
    cap drop `method'_prob_*
	cap drop `method'_los_*
    predictms, transmatrix(tmat) models(mod1 mod2 mod3_`method') prob los ///
	timevar(t) at1(age 60 sz2 0 sz3 0 nodes 0 pr_1 1 hormon 0) ci
    rename _prob_* `method'_prob_*
	rename _los_* `method'_los_*
\}

tw (line *_prob_at1_1_1 t, sort lc(red green orange blue)) ///
	(line *_prob_at1_1_2 t, sort lc(red green orange blue)) ///
	(line *_prob_at1_1_3 t, sort lc(red green orange blue)) ///
	, xtitle(Time (years)) ytitle(Probability) xlab(0(1)5) ylab(0(0.2)1, ///
	angle(h) format(\%4.1f)) legend(order(1 "Clock-Forward" 2 "Time At" ///
	3 "Time Since" 4 "Time At + Since") pos(9) ring(0) c(1)) ///
	title("Transition Probabilities") name(trans_models, replace) ///
	text(0.69 4 "Post Surgery", size(small)) ///
	text(0.15 4 "Relapse", size(small)) ///
	text(0.36 4 "Death", size(small)) ///
	graphregion(col(white))
graph export rott_trans_models.png, replace

tw (line *_los_at1_1_1 t, sort lc(red green orange blue)) ///
	(line *_los_at1_1_2 t, sort lc(red green orange blue)) ///
	(line *_los_at1_1_3 t, sort lc(red green orange blue)) ///
	, xtitle(Time (years)) ytitle(LOS (years)) xlab(0(1)5) ylab(0(1)5) ///
	legend(order(1 "Clock-Forward" 2 "Time At" ///
	3 "Time Since" 4 "Time At + Since") pos(9) ring(0) c(1)) ///
	title("LOS") name(los_models, replace) ///
	text(3.85 4.5 "Post Surgery", size(small)) ///
	text(0.49 4.5 "Relapse", size(small)) ///
	text(1 4.5 "Death", size(small)) ///
	graphregion(col(white))
graph export rott_los_models.png, replace
 
\end{stlog}

\end{document}